\documentclass[ aps, showpacs, showkeys, nofootinbib, floatfix, superscriptaddress]{revtex4}

\usepackage{amsfonts}

\usepackage{amssymb}
\usepackage{amsmath}
\usepackage{graphicx}
\usepackage{epstopdf}
\usepackage{soul}
\usepackage{xcolor}

\begin{document}

\title{Investigating the Physical Properties of Traversable Wormholes in the  Modified $f(R,T)$ Gravity }
 \author{Jianbo Lu}
  \email{lvjianbo819@163.com}
 \affiliation{Department of Physics, Liaoning Normal University, Dalian 116029, P. R. China}
  \author{Mou Xu}
 \affiliation{Department of Physics, Liaoning Normal University, Dalian 116029, P. R. China}
 \author{Jing Guo}
 \affiliation{Department of Physics, Liaoning Normal University, Dalian 116029, P. R. China}
 \author{Ruonan Li}
 \affiliation{Department of Physics, Liaoning Normal University, Dalian 116029, P. R. China}

\begin{abstract}
 Wormholes (WHs) are considered to be hypothetical tunnels connecting two distant regions of the universe or two different universes. In general relativity (GR), the formation of traversable WH requires the consideration of exotic matter that violates energy conditions (ECs). If the wormhole geometry can be described in modified gravitational theories without introducing exotic matter, it will be significant for studying these theories. In the paper, we analyze some physical properties of static traversable WH within the framework of $f(R,T)$ modified gravitational theory. Firstly, we explore the validity of the null, weak, dominant and strong energy conditions for wormhole matter for the considered $f(R,T)=R+\alpha R^2+\lambda T$ model. Research shows that it is possible to obtain traversable WH geometry without bring in exotic matter that violates the null energy condition (NEC) in the $f(R,T)$ theory. The violation of the dominant energy condition (DEC) in this model may be related to quantum fluctuations or indicates the existence of special matter that violates this EC within the wormhole. Moreover, it is found that in the $f(R,T)=R+\alpha R^2+\lambda T$ model, relative to the GR, the introduction of the geometric term $\alpha R^2$ has no remarkable impact on the wormhole matter components and their properties, while the appearance of the matter-geometry coupling term $\lambda T$ can resolve the question that WH matter violates the null, weak and strong energy condition in GR. Additionally, we investigate dependency of the valid NEC on model parameters and quantify the matter components within the wormhole using the ``volume integral quantifier". Lastly, based on the modified Tolman-Oppenheimer-Volkov equation, we find that the traversable WH in this theory is stable. On the other hand, we use the classical reconstruction technique to derive wormhole solution in $f(R,T)$ theory and discuss the corresponding ECs of matter.  It is found that all four ECs (NEC, WEC, SEC and DEC) of matter in the traversable wormholes are valid in this reconstructed $f(R,T)$ model, i.e we provide a wormhole solution without introducing the exotic matter and special matter in $f(R,T)$ theory. 

\end{abstract}


\maketitle

\section{$\text{Introduction}$}

 The modified theory of general relativity (GR) is one of the hot areas of physics, because it has important significance for explaining the accelerated expansion of the universe\cite{1}, dark matter and other problems. Over the past several decades, various modified theories have been established, with $f(R)$ theory \cite{2,3} receiving widespread attention as a kind of modified gravitational theory. In 2011, Harko and his collaborators proposed an extension of the $f(R)$ theory, termed as the $f(R,T)$ theory \cite{4}, by considering the coupling between geometry and matter. In this theory, $R$ represents the Ricci scalar, and $T$ denotes the trace of the energy-momentum tensor. And the theory has been extensively applied to several fields such as cosmology \cite{5,6,7,8,9,10,11,12,13,14}, black hole physics \cite{15,16,17,18}, and gravitational wave physics \cite{19}.etc.

The wormholes (WHs) have been studied as early as the 1930s. WHs are hypothetical space-time structures in the universe, which can be regarded as tunnels connecting two different universes. The area of minimum surface connecting two regions is called the ``throat" of the WH. In 1916, Flamm firstly introduced the concept of WHs \cite{20}. In 1935, Einstein and Rosen, while studying the gravitational field equations, hypothesized the existence of curved structure connecting two distinct regions of space-time namely ``Einstein-Rosen bridge" \cite{21}. However, WH solutions derived from the Einstein field equations require matter to violate the null energy condition (NEC). Decades later, Morris and Thorne proposed the concept of traversable WH and wrote its geometric structure in the form of the Morris-Thorne metric. In GR, the energy-momentum tensor of matter supporting such a structure also needs to violate the NEC near the WH throat \cite{22}. Typically, matter that violates the NEC is referred to as exotic matter \cite{23}, while matter violating other energy conditions (ECs) is called special matter (e.g., dark energy with negative pressure properties introduced in cosmology to explain the late-time accelerated expansion of the universe \cite{24,25}). In fact, researchers have found that traversable WH can still be formed without introducing exotic or special matter in the framework of modified gravitational theories, as these nonstandard wormhole geometries can be supported by the higher-order curvature term \cite{26}. Therefore, constructing traversable WH geometry in different modified theories is currently one of the research hotspots. To describe a static traversable WH, two functions must be inserted in the Morris-Thorne metric: the redshift function $\phi(r)$ and the shape function $b(r)$. Moreover, these two functions must satisfy the following conditions \cite{8,22,28,29,30,31,32,33,34,35,36,37,38,39,41}: (1) The radial coordinate $r$ is defined in the domain from the minimum radius value $r_0$ to $\infty$, i.e. $r\leq r_0 <\infty$, where $r_0$ describes the radius of the WH throat; (2) A flaring out condition of the throat is regarded as an important condition to have typical wormhole solutions, such that $\frac{b- b^{'}r}{b^2}>0$; (3) And at the throat $r=r_0=b(r_0)$, the relation $b^{'}(r_0)<1$ must be satisfied, where the prime denotes the derivative with respect to the radial coordinate $r$; (4) Another condition that the shape function needs to obey is: $1-\frac{b(r)}{r}>0$; (5) In order to ensure the absence of horizons and singularities, the redshift function $\phi(r)$ needs to be finite and non-null in the entire space-time. Although the existence of WHs have not yet been observed, important progress has been made in this field in recent years. For instance, in reference \cite{42}, the authors discussed the possibility of the wormhole in the Galactic halo, and the result indicated that the space-time of the Galactic halo could be described by a traversable wormhole geometry, which is consistent with the observed flat Galactic rotation curves.

In this paper, we study WH physics within the framework of $f(R,T)$ modified gravitational theory, checking the relevant physical properties of WH by selecting a model of the form $f(R,T)=f_1(R)+f_2(T)$. Specifically, we consider $f_1(R)=R+\alpha R^2$, $f_2(T)=\lambda T$. where $\alpha$ and $\lambda$ are two constants, and when the model parameters $\alpha=\lambda=0$, the theory reduces to GR. The model of $f_1(R)$ was first proposed by Starobinsky in an attempt to solve the inflation problem\cite{43}. In recent years, the Starobinsky models have played a vital role in many applications, such as analyzing matter density perturbations and the acceleration expansion of the universe during its later stages \cite{44,45,46}.

The classical reconstruction technique is usually utilized to derive the action of gravitational field in modified gravitational theories \cite{refr1,refr2,refr3,refr4,refr5}. One can note that using this technique to derive wormhole solution in $f(R,T)$ theory and discuss the corresponding energy conditions of matter has not been done so far. Here we try to construct an analytical function  for  theory and investigate its property of matter in wormhole.

The structure of the paper is as follows. In Section 2, we briefly introduce the $f(R,T)$ modified theory and provide an overview of the basic equations for WH physics within the framework of theory. In Section 3, we analyze the energy condition of WH matter in the theoretical model and other properties of traversable WH. Section 4 is the conclusion.

\section{$\text{ The Modified Gravitational Theory of $f(R,T)$ and Field Equations}$}

Assume the action of the modified theory has the following form \cite{4}:
\begin{eqnarray}
S=\frac{1}{16 \pi G} \int \sqrt{-{g}}[f(R,T)+16\pi G \mathcal{L}_{m}]d^{4}x.
\label{1}
\end{eqnarray}
where $\mathcal{L}_{m}$ is the Lagrangian density of matter, and we define the energy-momentum tensor of matter as
\begin{eqnarray}
T_{\mu \nu}=-\frac{2}{\sqrt{-g}} \frac{\delta\left(\sqrt{-g} \mathcal{L}_{m}\right)}{\delta g^{\mu \nu}}.
\label{2}
\end{eqnarray}
the trace of the energy-momentum tensor is represented $T=g^{\mu\nu}T_{\mu \nu}$, we define the variation of $T$  with respect to the metric tensor as
\begin{eqnarray}
\frac{\delta (g^{\alpha\beta}T_{\alpha\beta})}{\delta g^{\mu\nu}}{\delta{g^{\mu\nu}}}=T_{\mu\nu}+\theta_{\mu\nu}.\label{3}
\end{eqnarray}
where
\begin{eqnarray}
\theta_{\mu \nu}=g^{\alpha\beta}\frac{\delta T_{\alpha\beta}}{\delta g^{\mu\nu}}. \label{4}
\end{eqnarray}
Varying the action (1) gives the field equation
\begin{eqnarray}
R_{\mu\nu}f_R(R,T)-\frac{1}{2}g_{\mu\nu}f(R,T)+(g_{\mu\nu}\square -\nabla_\mu \nabla_\nu)f_R(R,T)=8\pi GT_{\mu\nu}-f_T(R,T)(T_{\mu\nu} + \theta_{\mu \nu}).\label{5}
\end{eqnarray}
where $f_{R}(R,T)=\frac{\partial f(R,T)}{\partial R}$, $f_{T}(R,T)=\frac{\partial f(R,T)}{\partial T}$, $\nabla_\mu$ represents the covariant derivative. And the covariant derivative of the energy-momentum tensor is given by the following formula \cite{48}:
\begin{eqnarray}
\nabla^{\mu} T_{\mu \nu}=\frac{f_{T}(R, T)}{8 \pi G-f_{T}(R, T)}\left[\left(T_{\mu \nu}+\theta_{\mu \nu}\right) \nabla^{\mu} \ln f_{T}(R, T)+\nabla^{\mu} \theta_{\mu \nu}-\frac{1}{2} g_{\mu \nu} \nabla^{\mu} T\right]. \label{6}
\end{eqnarray}
Unlike the case in GR where the energy momentum tensor is conserved, the energy momentum tensor is not conserved in $f(R,T)$ theory, i.e., $\nabla^{\mu} T_{\mu\nu}\neq0 $ . In the study of WH physics, the energy-momentum tensor of matter is typically considered to have an anisotropic distribution:
\begin{eqnarray}
T_{\mu\nu}=(\rho+P_t)u_\mu u_\nu-P_tg_{\mu\nu}+(P_r-P_t)\chi_\mu\chi_\nu.\label{7}
\end{eqnarray}
where $\rho$, $P_r$ and $P_t$ are the energy density, radial pressure and transverse pressure, respectively. Furthermore, we have: $u^\mu u_\mu=1$, $\chi^\mu \chi_\mu=-1$. If the Lagrangian of the matter field is written as $\mathcal{L}_{m}=-P $, therefore, $\theta_{\mu\nu}$ becomes
\begin{eqnarray}
\theta_{\mu\nu}=-2T_{\mu\nu}-Pg_{\mu\nu}.\label{8}
\end{eqnarray}
here $P=\frac{P_r+2P_t}{3}$ is the total pressure, then the trace of the energy-momentum tensor is $T=\rho-3P$. Consider the model $f(R,T)=R+\alpha R^2+\lambda T$, the field equation becomes
\begin{eqnarray}
G_{\mu\nu}=R_{\mu\nu}-\frac{1}{2} g_{\mu\nu} R=T^{eff}_{\mu\nu}.\label{9}
\end{eqnarray}
\begin{eqnarray}
T^{eff}_{\mu\nu}=\frac{1}{1+2\alpha R} \{[(8\pi+\lambda)T_{\mu\nu}+\lambda Pg_{\mu\nu}]+[\frac{1}{2}[R+\alpha R^2+\lambda T-R(1+2\alpha R)]g_{\mu\nu}]-[(g_{\mu\nu}\square -\nabla_\mu \nabla_\nu)f_R]\}.\label{10}
\end{eqnarray}
Clearly, the effective energy-momentum tensor is conserved by virtue of the Bianchi identity. In other words, the covariant derivative of $T^{eff}_{\mu\nu}$ equals to zero, i.e., $\nabla^\mu T^{eff}_{\mu\nu}=0$. In $f(R,T)$ theory, due to the coupling between geometry and matter, the four divergence of the energy-momentum tensor does not vanish. Thus, the conservation equation is usually rewritten by using the effective energy-momentum tensor.

To research the WH geometry under the modified gravitational framework, we start from the following Morris-Thorne static spherically symmetric metric \cite{22}:
\begin{eqnarray}
d S^{2}=e^{2\phi(r)}dt^{2}-\frac{d r^2}{1-\frac{b(r)}{r}}-r^{2}(d\theta^2+\sin^2\theta d\phi^2).\label{11}
\end{eqnarray}
where $\phi(r)$ and $b(r)$ are the redshift function and shape function. Throughout in this paper, we take the metric signature as (+, -, -, -). Obviously, this convention used for the metric signature is consistent with Refs.\cite{30,mt1,mt3,mt4,mt5,mt6,mt7}. For simplicity, we take the redshift function as a constant.

\section{$\text{Physical Properties of Traversable Wormhole in $f(R,T)=R+\alpha R^2+\lambda T$} theory $}

Under the static spherically symmetric metric, in the $f(R,T)=R+\alpha R^2+\lambda T$ model, the components of the field equations are represented as:
\begin{eqnarray}
\frac{b^{\prime}}{r^{2}}=\frac{1}{1+2 \alpha R}\left[\left(8 \pi+\frac{3 \lambda}{2}\right) \rho-\frac{\lambda}{6}\left(p_{r}+2 p_{t}\right)-\frac{2 \alpha b^{\prime 2}}{r^{4}}\right].\label{12}
\end{eqnarray}
\begin{eqnarray}
\frac{b}{r^{3}}=\frac{1}{1+2 \alpha R}\left[-\left(8 \pi+\frac{7 \lambda}{6}\right) p_{r}+\frac{\lambda}{2}\left(\rho-\frac{2}{3} p_{t}\right)-\frac{2 \alpha b^{\prime 2}}{r^{4}}\right].\label{13}
\end{eqnarray}
\begin{eqnarray}
\frac{b^{\prime} r-b}{2 r^{3}}=\frac{1}{1+2 \alpha R}\left[-\left(8 \pi+\frac{4 \lambda}{3}\right) p_{t}+\frac{\lambda}{2}\left(\rho-\frac{1}{3} p_{r}\right)-\frac{2 \alpha b^{\prime 2}}{r^{4}}\right].\label{14}
\end{eqnarray}
with $R=-\frac{2b^{'}}{r^2}$, Combining equations (12)-(14), we obtain
\begin{eqnarray}
\rho=\frac{b^{\prime}\left[\lambda\left(2 r^{2}-5 \alpha b^{\prime}\right)+12 \pi\left(r^{2}-2 \alpha b^{\prime}\right)\right]}{3(\lambda+4 \pi)(\lambda+8 \pi) r^{4}} .\label{15}
\end{eqnarray}
\begin{eqnarray}
p_{r}=-\frac{-12 \alpha \lambda b b^{\prime}-48 \pi \alpha b b^{\prime}-\lambda r^{3} b^{\prime}+7 \alpha \lambda r b^{\prime 2}+24 \pi \alpha r b^{\prime 2}+3 b \lambda r^{2}+12 \pi b r^{2}}{3(\lambda+4 \pi)(\lambda+8 \pi) r^{5}}.\label{16}
\end{eqnarray}
\begin{eqnarray}
p_{t}=-\frac{12 \alpha \lambda b b^{\prime}+48 \pi \alpha b b^{\prime}+\lambda r^{3} b^{\prime}+12 \pi r^{3} b^{\prime}+2 \alpha \lambda r b^{\prime 2}-3 b \lambda r^{2}-12 \pi b r^{2}}{6(\lambda+4 \pi)(\lambda+8 \pi) r^{5}}.\label{17}
\end{eqnarray}
Equations (\ref{15}) - (\ref{17}) determines the relationship between the gravitational field and the matter field of a traversable wormhole in the $f(R,T)=R+\alpha R^2+\lambda T$ modified theory.

Several viable shape functions have been widely proposed and applied to WH physics. For example, Lob and Oliveira \cite{49} employed shape functions $b(r)=\frac{r_0^2}{r}$ and $b(r)=\sqrt{r r_0}$ to analyze traversable WH in $f(R)$ theory; Rahaman et al.\cite{50} utilized shape functions $b(r)=r_0+\rho_0r^3_0\ln{(\frac{r_0}{r})}$ and $b(r)=r_0+\gamma r_0(1-\frac{r_0}{r})$ to investigate Finslerian wormhole models (where $\rho_0$ and $\gamma$ are arbitrary constants  less than 1). We choose the popular power-law form of the shape function $b(r)=\frac{r_0^2}{r}$ to study the properties of the wormhole in the framework of the $f(R,T)=R+\alpha R^2+\lambda T$ theoretical model. For the discussion by taking some other shape functions to investigate WH this $f(R,T)$ model, one can see references \cite{mt3,mt7}.

\subsection{$\text{Energy conditions}$}

In the study of modified gravitational theories, the validity of energy conditions of matter are often the key issue. And the ECs are the necessary conditions to explain singularity theorem \cite{51}. Moreover, the ECs not only aid in analyzing the entire space-time structure without precise solutions to Einstein's equations but also play a crucial role in the investigation of wormhole solutions in the context of modified gravity theories. The common popular basic energy conditions (the null, weak, dominant and strong energy condition) originate from the Raychaudhuri equation \cite{52}, which play crucial roles in describing the attractive properties of gravity and positive energy density. Specifically, the expressions for various ECs can be written as \cite{53}:

$\bullet $ Null Energy Condition (NEC): $\rho+P_r\geq0$, $\rho+P_t\geq0$;

$\bullet $ Weak Energy Condition (WEC): $\rho\geq0$, $\rho+P_r\geq0$, $\rho+P_t\geq0$;

$\bullet $ Dominant Energy Condition (DEC): $\rho\geq0$, $\rho-\left|P_r\right|\geq0$, $\rho-\left|P_t\right|\geq0$;

$\bullet $ Strong Energy Condition (SEC): $\rho+P_r\geq0$, $\rho+P_t\geq0$, $\rho+P_r+2P_t\geq0$;

From the above expressions, it is easy to see that if the NEC is violated, which implies that all the ECs mentioned above should be broken. We focus on exploring the realization of traversable WH geometry in the $f(R,T)$ theory by avoiding the introduction of exotic matter (violating NEC). substituting $b(r)=\frac{r_0^2}{r}$ into equations (15)-(17), we can obtain the energy density, radial pressure and transverse pressure, respectively (with wormhole throat radius $r_0=1$):
\begin{eqnarray}
\rho=-\frac{r_{0}^{2}\left[2 r^{4}(6 \pi+\lambda)+\alpha(24 \pi+5 \lambda) r_{0}^{2}\right]}{3 r^{8}(4 \pi+\lambda)(8 \pi+\lambda)}.\label{18}
\end{eqnarray}
\begin{eqnarray}
p_{r}=-\frac{r_{0}^{2}\left[4 r^{4}(3 \pi+\lambda)+\alpha(72 \pi+19 \lambda) r_{0}^{2}\right]}{3 r^{8}(4 \pi+\lambda)(8 \pi+\lambda)}.\label{19}
\end{eqnarray}
\begin{eqnarray}
p_{t}=\frac{2 r^{4}(6 \pi+\lambda) r_{0}^{2}+\alpha(24 \pi+5 \lambda) r_{0}^{4}}{3 r^{8}(4 \pi+\lambda)(8 \pi+\lambda)}.\label{20}
\end{eqnarray}

\begin{figure}[ht]
  \includegraphics[width=6.5cm]{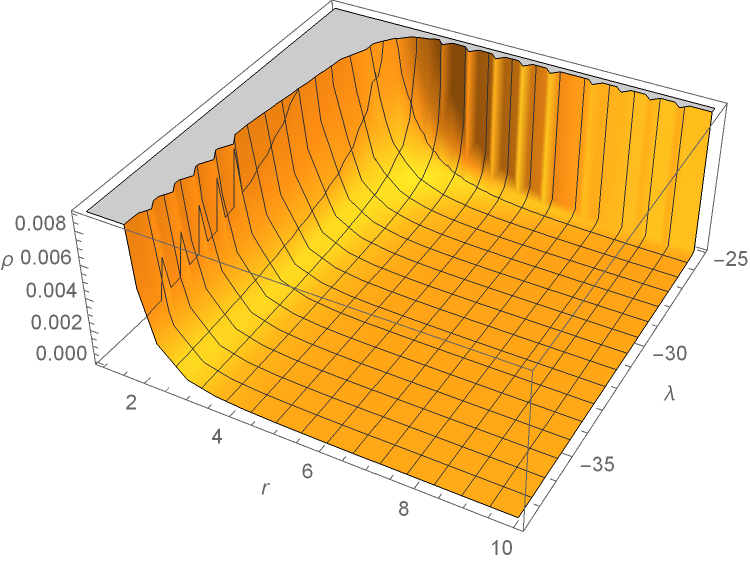}
  \includegraphics[width=6.5cm]{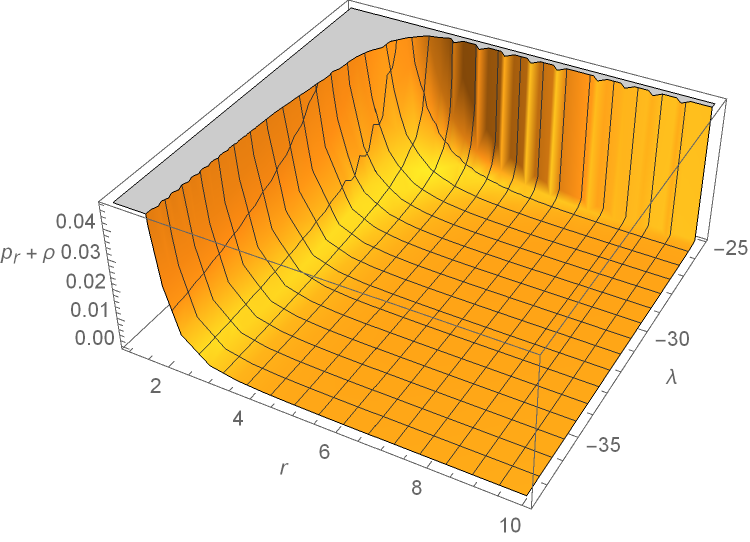}\\
    \includegraphics[width=6.5cm]{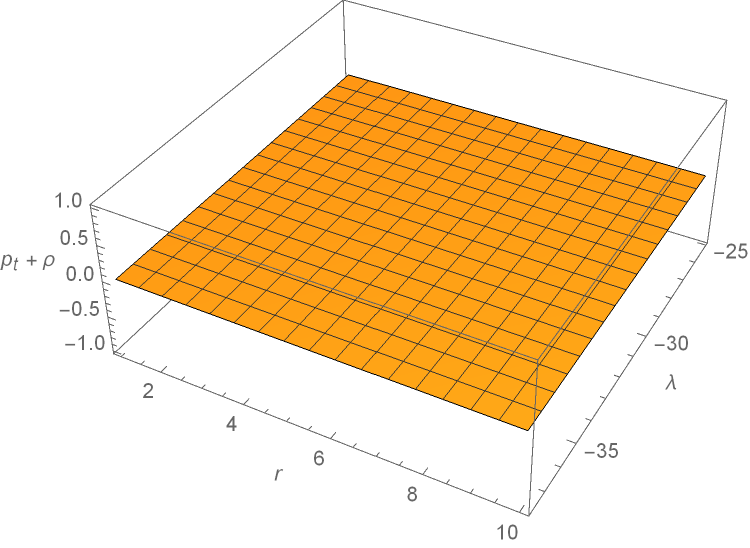}
  \includegraphics[width=6.5cm]{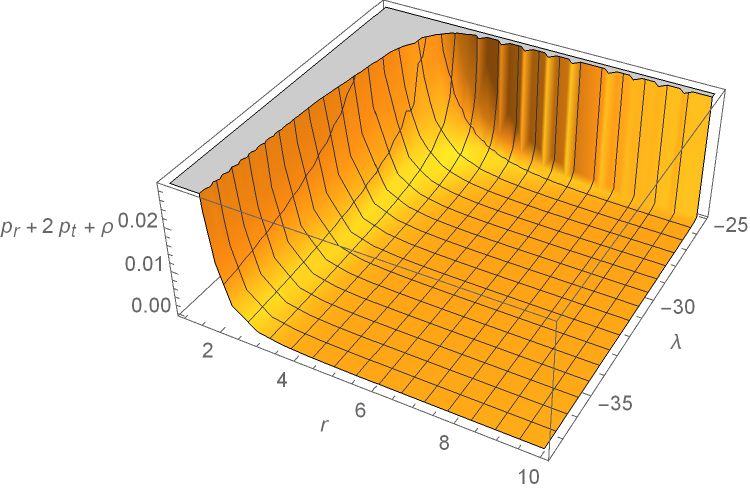}\\
  \includegraphics[width=6.5cm]{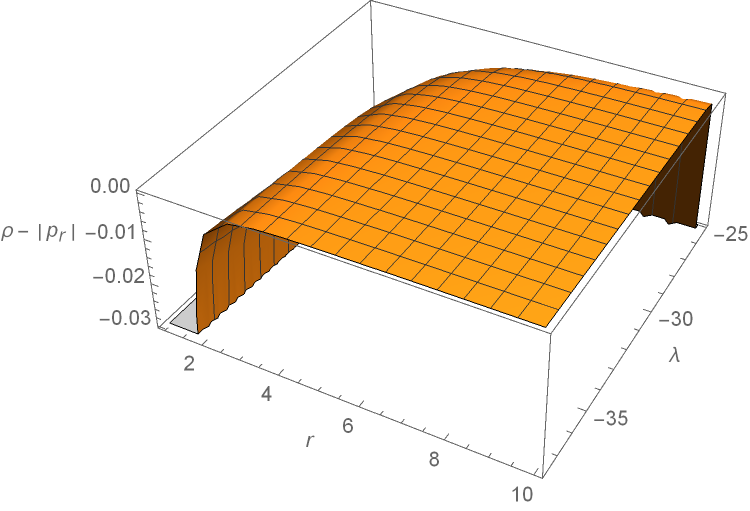}
  \includegraphics[width=6.5cm]{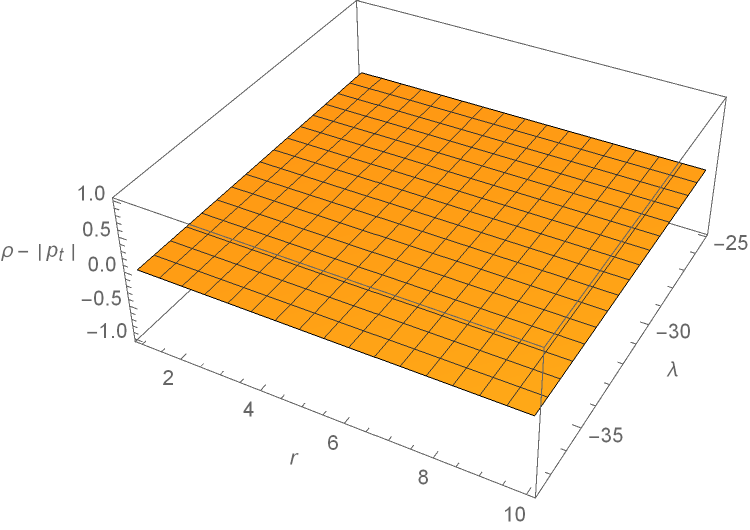}\\
  \caption{ Pictures of the relative expressions of energy conditions with respect to parameters $r$ and $\lambda$ in the $f(R,T)$ wormhole, where the model parameter value $\alpha=5$ is taken.}\label{f1}
\end{figure}

In the early 1980s, Starobinsky \cite{43} proposed the model: $f(R)=R+\alpha R^2$ $(\alpha>0)$ to solve the problem of cosmic inflation. For the $f(R,T)=R+\alpha R^2+\lambda T$ gravitational model explored in this work, we also regard as the model parameter to meet the relationship: $\alpha>0$, which allows for cosmic inflation when parameter $\lambda=0$. From equation (\ref{18}), we can easily find that to guarantee the energy density is always positive, another parameter needs to satisfy $\lambda<-8\pi$. Next, based on the derived expressions (\ref{18})-(\ref{20}), we plot three-dimensional pictures for the energy conditions of the wormhole matter relative $r$ in the $f(R,T)$ theory. In Figure \ref{f1}, we consider taking the value of model parameter as: $\alpha=5$; the value of $\lambda$ is limited to: $-12\pi<\lambda<-8\pi$. The $f(R,T)$ theory takes the value of the above parameters, the matter near the WH throat satisfies NEC, WEC and SEC, but the radial DEC is not satisfied. The superluminal phenomenon implied by the violation of DEC \cite{54,55} may be related to the quantum fluctuations of WH matter \cite{56}, or it may suggest the existence of special matter near the WH throat. 

\begin{figure}[ht]
\includegraphics[width=7.5cm, height=5cm]{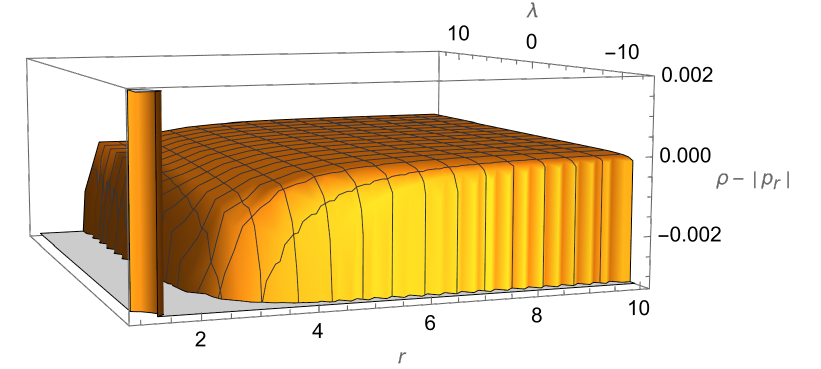}
\includegraphics[width=7.5cm, height=5cm]{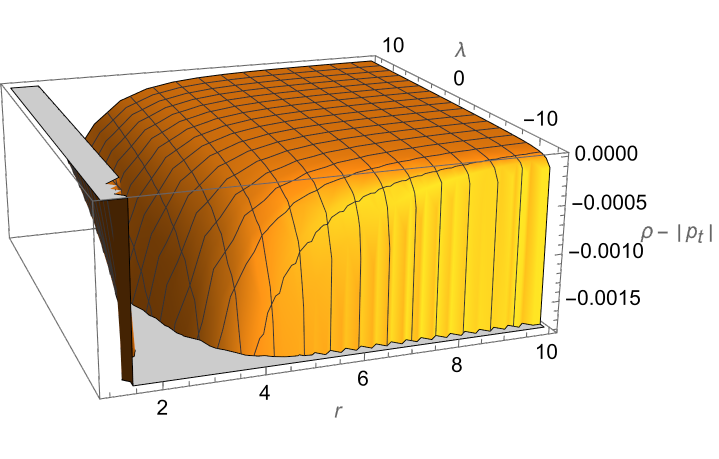}\\
\includegraphics[width=7.5cm, height=5cm]{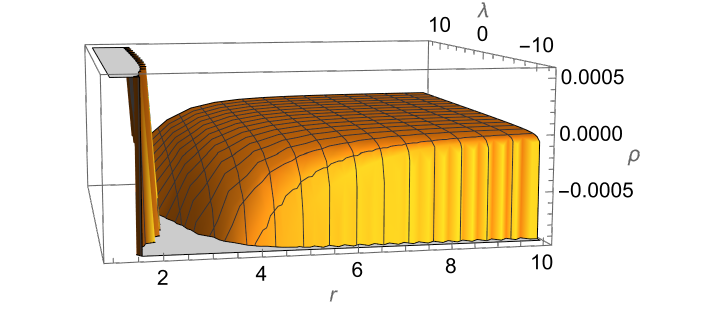}
\includegraphics[width=7.5cm, height=5cm]{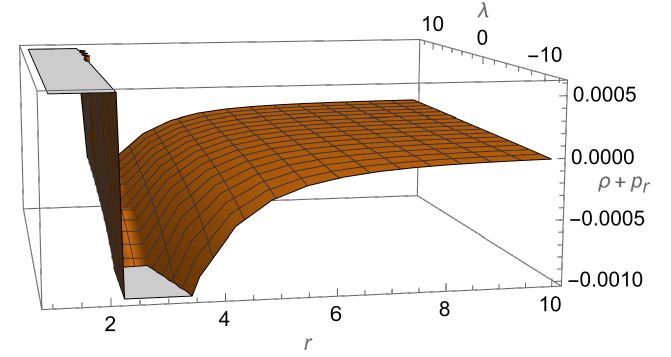}\\
\includegraphics[width=7.5cm, height=5cm]{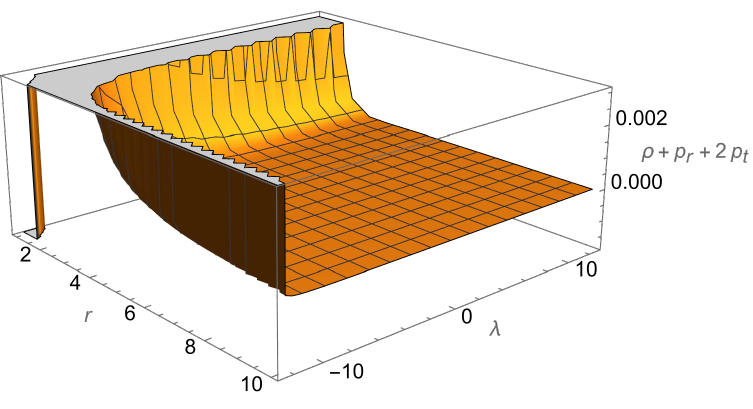}
\includegraphics[width=7.5cm, height=5cm]{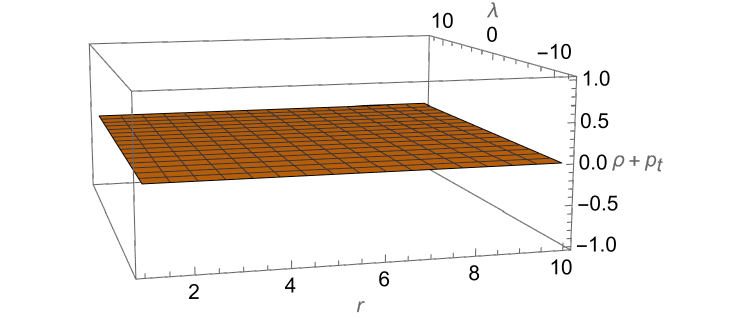}\\
\caption{Pictures of the relative expressions of energy conditions with respect to parameters $r$ and $\lambda$ in the $f(R, T)$ wormhole, where the model parameter value $\alpha=-5$ is taken.}\label{add-f1}
\end{figure}

To consider the impact of different $\alpha$ values on energy conditions, we also explore the validity of energy conditions for matter in the $f (R, T)$ theoretical model when $\alpha<0$. For the $f (R, T)$ model selected in this article, the $\lambda$ value is limited to: $-4\pi<\lambda<4\pi$. In Figure \ref{add-f1}, we draw pictures showing the changes in the expression of  material energy conditions relative to $r$ and $\lambda$ when the model parameter is set to $\alpha=-5$. From the figure, it can be seen that near the wormhole throat, the matter satisfies NEC, WEC, and SEC, while the tangential DEC  is violated. For the matter far from the throat of wormhole, none of the four energy conditions are met. Obviously, all four types of the pointwise energy conditions in this $f(R,T)$ model corresponding to $\alpha=-5$ are not satisfied. Therefore, in the following text, we will only consider the case when $\alpha=5$.

\begin{figure}[ht]
\includegraphics[width=7.5cm]{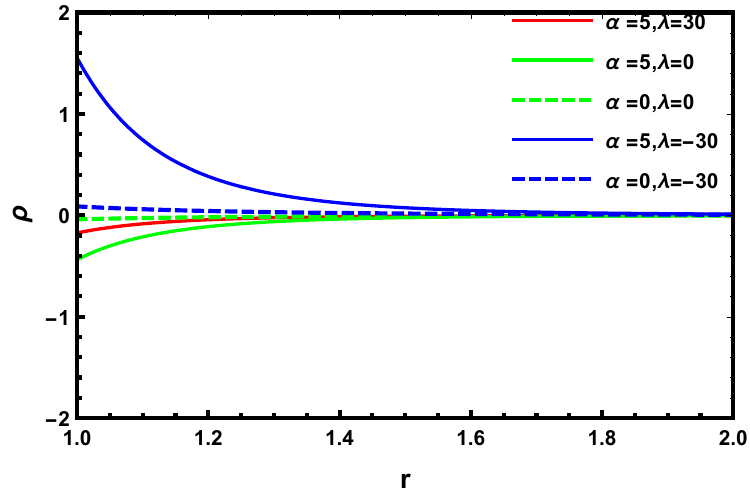}
  \includegraphics[width=7.5cm]{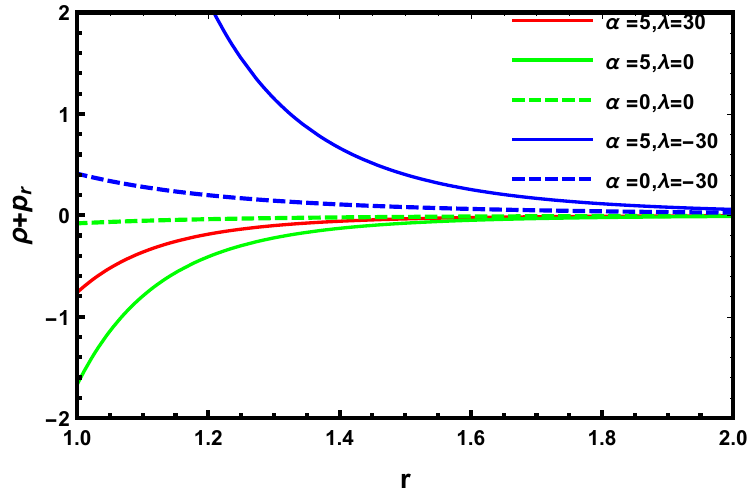}\\
    \includegraphics[width=7.5cm]{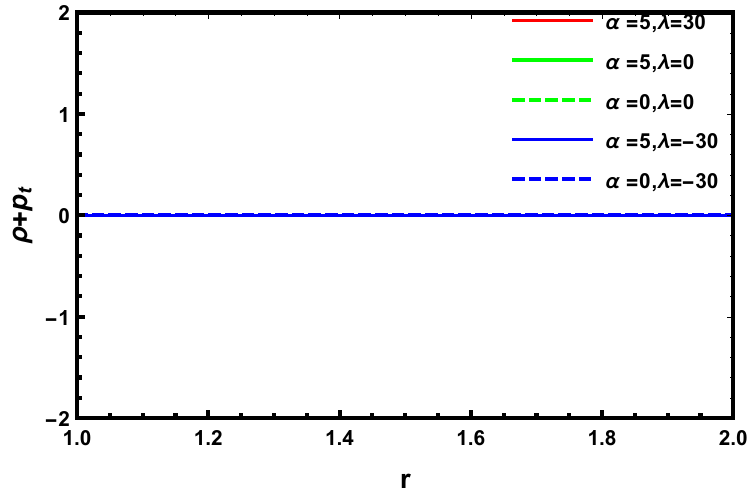}
  \includegraphics[width=7.5cm]{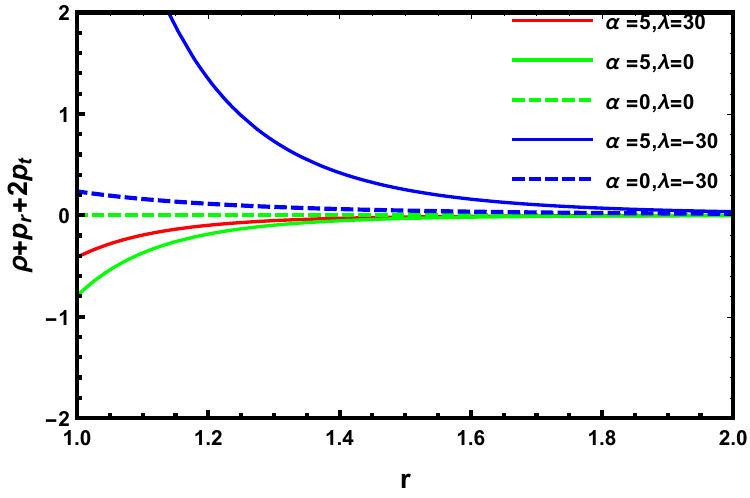}\\
  \includegraphics[width=7.5cm]{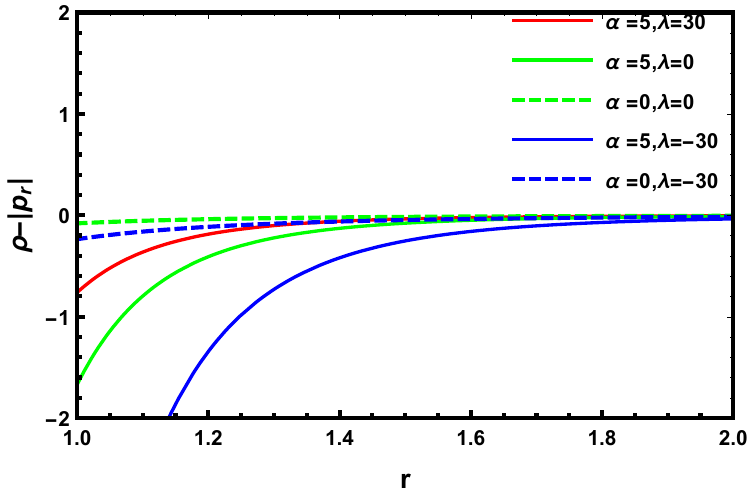}
  \includegraphics[width=7.5cm]{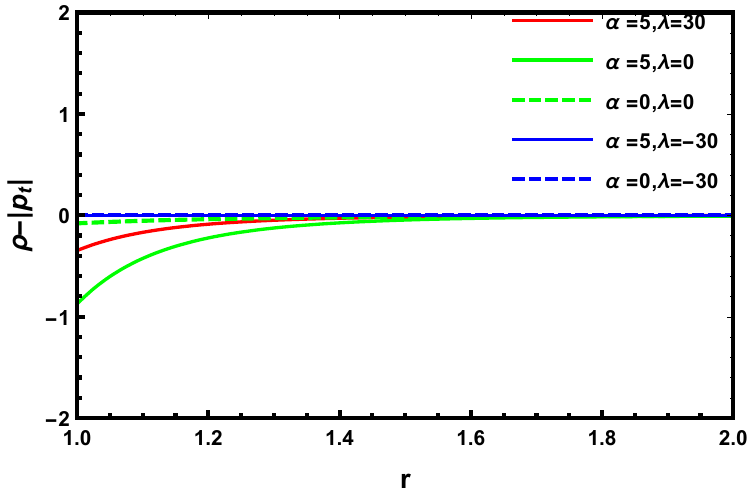}\\
  \caption{ Pictures of the relevant expressions of energy conditions for matter in the wormhole under $f(R,T)$ theory, with taking the different values of model parameters $ \alpha$ and $\lambda$.}\label{f2}

\end{figure}

To visually demonstrate the variations of energy conditions expressions with respect to the model parameter values, we draw two-dimensional pictures in Figure \ref{2} (considering $\alpha=5$, $\lambda=0$, $\pm30$). In order to investigate  the influence of each term in the $f(R,T)=R+\alpha R^2+\lambda T$ model on the ECs, we discuss the variations of the ECs of wormhole matter in their associated models. (1) When the parameter $\lambda=0$, the $f(R,T)$ theory recover to the $f(R)=R+\alpha R^2$ model. As seen from Figure \ref{f2}, with model parameter $\alpha=5$ the matter near the WH throat does not satisfy the four ECs mentioned above, implying that exotic matter is required to maintain the stability of the WH; (2) For $\alpha=0$ , the model degenerates to: $f(R,T)=R+\lambda T$, which a modified model relative to GR with a matter-geometry coupling term. As an example, we also plot the two-dimensional graphics of the relevant energy conditions expressions with respective to the variable $r$ when $\alpha=0$, $\lambda=-30$ (as shown in Figure \ref{f2}). From the figure, it is indicated that the radial DEC of wormhole matter is violated. Comparing the results of the three models, we find that, relative to GR theory, the insertion of the matter-geometry coupling term $\lambda T$ in the $f(R,T)=R+\lambda T$ theoretical model has a more noticeable impact on the properties of wormhole matter. It can resolve the problem of WH matter violating the NEC, WEC and SEC in GR, leading to changes the composition and related properties of WH matter; Nevertheless, the introduction(relative to GR) of the $\alpha R^2$ geometric term in the $f(R)=R+\alpha R^2$ model cannot fulfill the requirements of the NEC, WEC and SEC for matter.


\subsection{$\text{ Volume integral quantifier}$}

Researches on the energy conditions of matter can usually be considered by the two following approaches: pointwise energy conditions and averaged energy conditions. In GR, the fundamental requirement for constructing a WH geometry is the matter violates the NEC \cite{22}. In fact, in traversable WH, matter violates all pointwise energy conditions and averaged energy conditions. The pointwise ECs refer to the conditions satisfied by the energy-momentum tensor at each point in space-time \cite{57}. The averaged energy conditions permit localized violations of the energy conditions, as long on average the energy conditions hold when integrated along timelike or null geodesics \cite{58}. Obviously, the averaged ECs are weaker than their corresponding pointwise ECs, since it is allowable that the localized violations for the averaged ECs \cite{58}. The most useful of the averaged ECs is typically the averaged NEC \cite{viq-0}, which can be  described as \cite{viq-1}.
\begin{eqnarray}
\int T_{\mu\nu}k^{\mu}k^{\nu}d\lambda \ge 0
\label{A16up}
\end{eqnarray}
where $k^{\mu}$ is the tangent vector along a null geodesic and $\lambda$ is the affine parameter labeling points on the geodesic.

It is shown in Refs.\cite{59} that the averaged null energy condition involve a line integral with dimensions (mass)/(area) rather than a volume integral, which does not provide useful information about the "total amount" of matter violating ECs. So, Visser et al.\cite{59,viq-1} proposed the "volume integral quantifier" (VIQ) for quantifying the total amount of matter violating null energy condition, i.e. the amount of "exotic matter" could be quantified by the extent of the negative value for VIQ. Furthermore, Refs.\cite{59,viq-1}constructed traversable WHs with arbitrarily small quantities of averaged NEC-violating matter, and indicated that quantum physics is known to lead to small violations of the averaged NEC.

Usually, matter that violates the NEC is referred to as exotic matter. To investigate the amount of exotic matter near the throat of the wormhole, one can calculate the following VIQ:
\begin{eqnarray}
I_v=\oint{(\rho+P_r)}dV =8\pi\int\nolimits^{\infty}_{r_0}(\rho+P_r)r^2 dr.\label{21}
\end{eqnarray}
We bring the formulas (\ref{18}) - (\ref{19}) into the above equation, and then the value of VIQ is
\begin{eqnarray}
I_v=[-\frac{16\pi(-\frac{1}{r}-\frac{4\alpha}{5r^5})}{8\pi+\lambda}]^{\infty}_{r_0}.\label{22}
\end{eqnarray}

\begin{figure}[ht]
  \includegraphics[width=7.5cm]{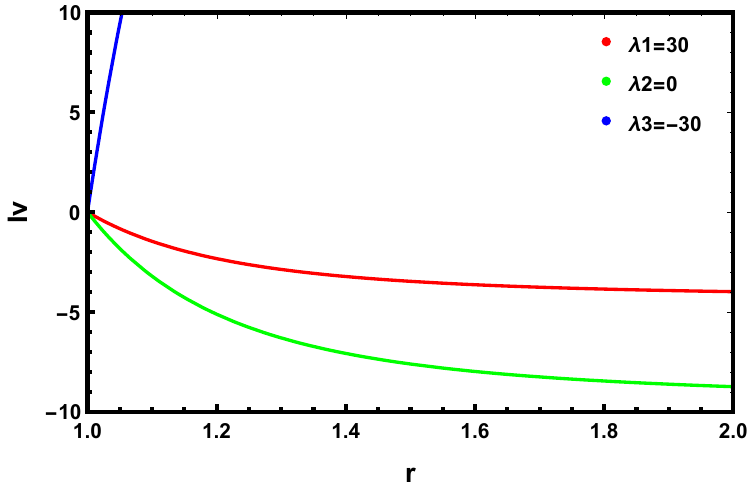}\\
  \caption{ Pictures of $I_v$ relative to $r$ in the  wormhole under $f(R,T)$ theory, where the different values of model parameters $\lambda$ are taken. }\label{f3}
\end{figure}

Using expression (\ref{22}), we can illustrate the variation of VIQ with respect to $r$ in Figure \ref{f3}. Here, the values of the model parameters are chosen as: $\alpha=5$, $\lambda=0$,
$\pm30$. From the picture, we observe that: when $r\rightarrow1$, all three cases have $I_v\rightarrow0$. Moreover, for $r>1$, when $\lambda=-30$, we have $I_v>0$, implying the realization of traversable WH in the $f(R,T)$ theory does not require the introduction of exotic matter. However, for the parameter value $\lambda=0$ and $30$, we have gotten $I_v<0$, indicating the presence of exotic matter at the throat of the WH.

\subsection{$\text{ Tolman-Openeheimer-Volkoff equation}$}

\begin{figure}[ht]
  \includegraphics[width=9.5cm]{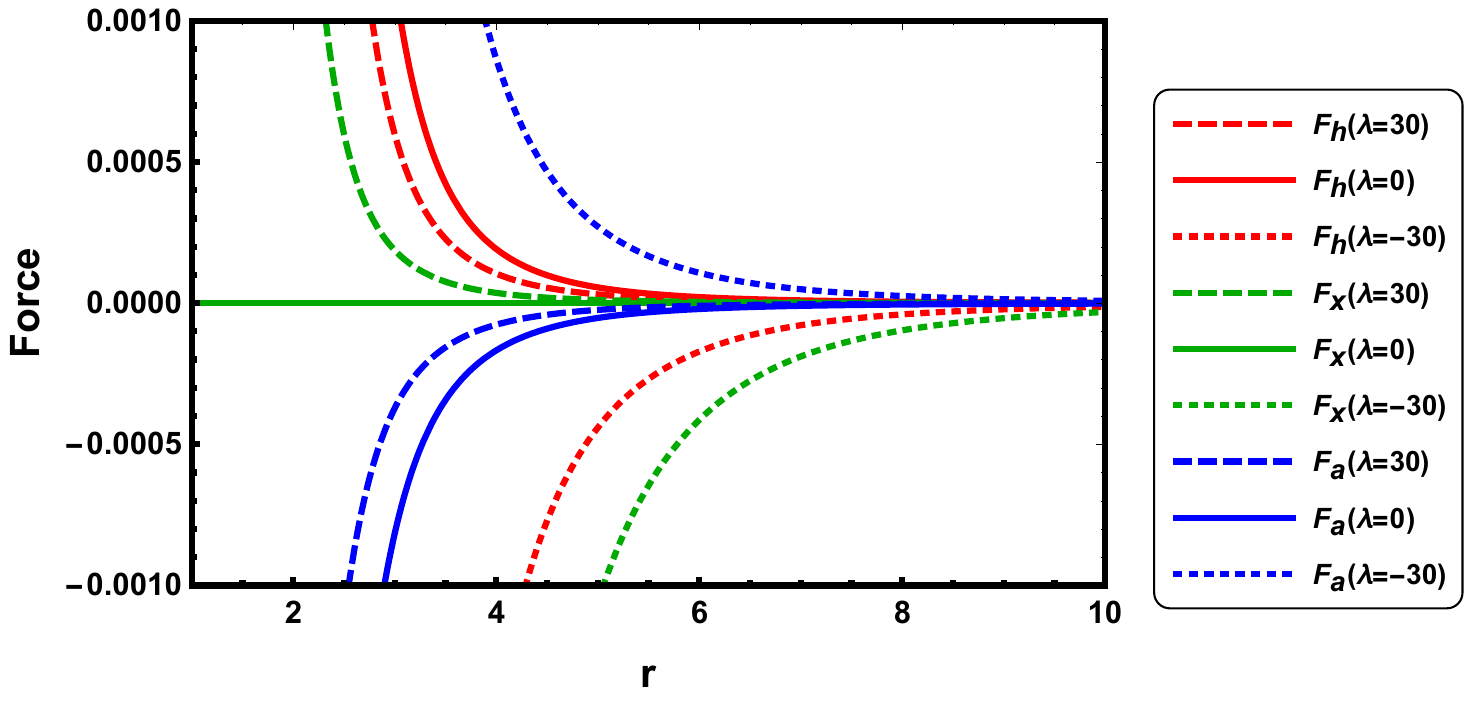}\\
  \caption{ Under the theory of $f(R,T)$, pictures of $F_h$, $F_a$ and $F_x$ relative to $r$ with taking the different value of model parameter $\lambda$. }\label{f4}
\end{figure}

All compact stars are believed to be in a state of equilibrium. Such equilibrium can be described  by using the modified Tolman-Oppenheimer-Volkoff (abbreviated as TOV) equation \cite{60}. This subsection discusses the effects of various forces on the current gravitational system. In the $f(R,T)$ theory, we derive the modified TOV equation as follows:
\begin{eqnarray}
\frac{d p_{r}}{d r}+\phi^{\prime}(\rho+p_{r})+\frac{2}{r}(p_{r}-p_{t})=-\frac{\lambda}{6(8 \pi+\lambda)}[\frac{d p_{r}}{d r}+\frac{d}{d r}(3 \rho-2 p_{t})]+\phi^{\prime}[(2+\frac{\lambda}{6(8 \pi+\lambda)}) \rho-\frac{\lambda}{3(8 \pi+\lambda)}(p_{r}+2 p_{t})]. \label{23}
\end{eqnarray}
When $\phi=0$, The above equation is changed to
\begin{eqnarray}
\frac{d p_{r}}{d r}+\frac{2}{r}(p_{r}-p_{t})=-\frac{\lambda}{6(8 \pi+\lambda)}[\frac{d p_{r}}{d r}+\frac{d}{d r}(3 \rho-2 p_{t})]. \label{24}
\end{eqnarray}
One can utilize $F_h$, $F_a$ and $F_x$ to rewrite the balance equation (\ref{24}). Concretely,   $F_{h}$ stands for the pressure gradient (or the hydrostatic force),   $F_{a}$ denotes the anisotropic force due to the anisotropic nature of matter, and $F_{x} $represents the the extra force from the discontinuous energy-momentum tensor in $f(R,T)$ gravity \cite{mt6,tov2,tov3,tov4}. Bringing the formulas (\ref{18}) -(\ref{20}) into (\ref{24}), Then we can exhibit $F_h $, $F_a$ and $F_x$ as follows:
\begin{eqnarray}
F_h=\frac{d p_{r}}{d r}=-\frac{16(3 \pi+\lambda)}{3 r^{5}(4 \pi+\lambda)(8 \pi+\lambda)}+\frac{8\left[4 r^{4}(3 \pi+\lambda)+\alpha(72 \pi+19 \lambda)\right]}{3 r^{9}(4 \pi+\lambda)(8 \pi+\lambda)}. \label{25}
\end{eqnarray}
\begin{eqnarray}
F_a=\frac{2}{r}\left(p_{r}-p_{t}\right)=-\frac{4\left(r^{4}+4 \alpha\right)}{r^{9}(8 \pi+\lambda)}. \label{26}
\end{eqnarray}
\begin{eqnarray}
F_x=\frac{\lambda}{6(8 \pi+\lambda)}\left[\frac{d p_{r}}{d r}+\frac{d}{d r}\left(3 \rho-2 p_{t}\right)\right]=\frac{4 \lambda\left(36 \pi r^{4}+192 \pi \alpha+7 r^{4} \lambda+44 \alpha \lambda\right)}{9 r^{9}(4 \pi+\lambda)(8 \pi+\lambda)^{2}}. \label{27}
\end{eqnarray}
 Based on the equations (\ref{25})-(\ref{27}), we draw the variations of $F_h $, $F_a$ and $F_x$ with respect to $r$ in the $f(R,T)$ gravity, with the model parameter values set as: $\alpha=5$, $\lambda=0$, $\pm30$. We can observe that the current gravitational system is in equilibrium state, as show in Figure \ref{f4}. When the model parameter is valued at $\lambda=30$, the combined effect of $F_h $ and $F_x$ is counterbalanced by $F_a$, and the force $F_a$ is attractive; On the other hand, when $\lambda=-30$, $F_a$ manifests repulsive; For $\lambda=0$, the effect of $F_h$ is counterbalanced by $F_a$ and the model $f(R)=R+\alpha R^2$ is retrieved.

\section{$\text{Validity of material ECs for WH solution given by the classical reconstruction technique in $f (R, T)$ theory}$}

\begin{figure}[ht]
\includegraphics[width=7.5cm, height=5cm]{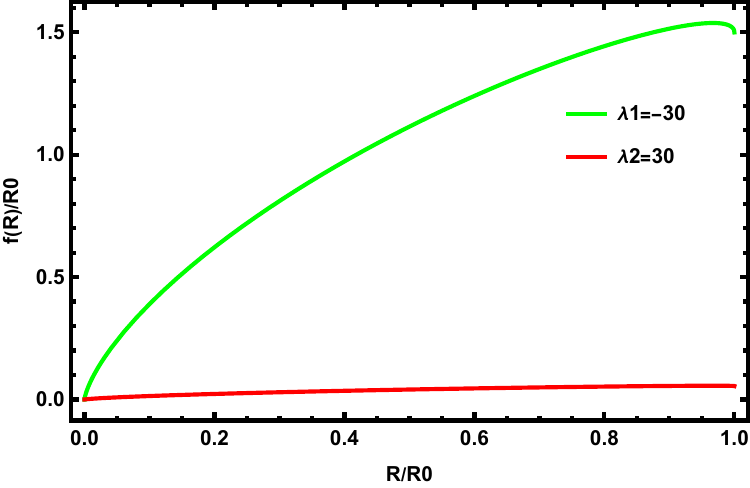}
\caption{The pictures of $f(R)$ in $f(R,T)$ theory for the used shape function: $b(r)=\frac{r^2_0}{r}$.}\label{add-ff1}
\end{figure}

Obviously, it is very interesting if one can obtain a WH solution without introducing the exotic matter and special matter in the modified gravitational theory. One can note that using the classical reconstruction technique to derive wormhole solution in $f(R,T)$ theory and discuss the energy conditions of matter in this solution has not been done so far. In this section, we also investigate the ECs of matter in $f (R, T)$ theory with the help of wormhole solution given by using the classical reconstruction technique. Consider the model $f(R,T)=f(R)+f(T)$ with $f(T)=\lambda T$, then the field equation in $f(R,T)$ theory becomes
\begin{eqnarray}
R_{\mu\nu}-\frac{1}{2}g_{\mu\nu}R=\frac{1}{f_R}[(8\pi+\lambda)T_{\mu\nu}+\lambda Pg_{\mu\nu}]+\frac{1}{f_R}[\frac{1}{2}(f(R)+\lambda T-Rf_R)]g_{\mu\nu}-\frac{1}{f_R}[(g_{\mu\nu}\square-\nabla_\mu\nabla_\nu)f_R].
\label{A1}
\end{eqnarray}
Taking the trace of gravitational field equation in $f(R,T)$ theory, we receive
\begin{eqnarray}
T=\frac{Rf_R+3\square f_R-4\lambda P-2f(R)}{8\pi+3\lambda}.
\label{A2}
\end{eqnarray}
Combining equation (\ref{A2}) with equation (\ref{A1}), we have
\begin{eqnarray}
R_{\mu\nu}-\frac{1}{2}g_{\mu\nu}R=T^{eff}_{\mu\nu}
\label{A3}
\end{eqnarray}
with
\begin{eqnarray}
T^{eff}_{\mu\nu}=\frac{1}{f_R}[(8\pi+\lambda)T_{\mu\nu}+\lambda Pg_{\mu\nu}]+\frac{1}{f_R}[-\frac{1}{4}[(8\pi+\lambda)T+4\lambda P]-\frac{1}{4}Rf_R+\frac{3}{4}\square f_R]g_{\mu\nu}-\frac{1}{f_R}[(g_{\mu\nu}\square-\nabla_{\mu}\nabla_{\nu})f_R].
\label{A4}
\end{eqnarray}
Using the MT metric and equation (\ref{A4}), we can derive to give the following equations
\begin{eqnarray}
\frac{b'}{r^2}=-\frac{1}{3}(p_r+2p_t)+\frac{1}{8\pi+\lambda}[\frac{4b'}{3r^2}f_R-\frac{1}{3}\square f_R]
\label{A6}
\end{eqnarray}
\begin{eqnarray}
\frac{b}{r^3}=\frac{1}{3}(-\rho+2p_t)+\frac{1}{8\pi+\lambda}[\frac{2(b'r-2b)}{3r^3}f_R-\frac{1}{3}\square f_R+\frac{4}{3}(1-\frac{b}{r})[f''_R+\frac{b'r-b}{2r^2(\frac{b}{r}-1)}f'_R]]
\label{A7}
\end{eqnarray}
\begin{eqnarray}
\frac{b'r-b}{2r^3}=\frac{1}{2}(-\rho+p_r)+\frac{1}{8\pi+\lambda}[\frac{2b'r-b}{r^3}f_R+\frac{1}{2}\square f_R-\frac{2}{f_Rr^2}(b-r)f'_R]
\label{A8}
\end{eqnarray}
where $\square f_R=-(1-\frac{b}{r})[f''_R-\frac{b-b'r}{2r(b-r)}f'_R]$, and the prime represents a derivative relative to the radial coordinate $r$. It is observed that the property of matter (including its energy density and pressure), which associated with $b(r)$ and $f_R(r)$, can be deduced  from equations (\ref{A6})–(\ref{A8}). Several methods have been explored to solve this kinds of equations in wormhole physics of modified gravitational theories. Following the strategy shown in Refs.cite{49,Added-44}, next we solve equations (\ref{A6})-(\ref{A8}) to give the traversable wormhole solution for $f(R,T)$ theory, by taking the following two equations of state. One is that we take into account the traceless stress-energy tensor, $T=-\rho+p_r+2p_t=0$, which is often associated to the Casimir effect, with a massless field \cite{49}. For this case, the $f(R,T)$ model will reduce to the $f(R)$ theory, and the corresponding  research has been done in reference \cite{49}. In reference \cite{49}, the traversable wormhole solution has been proposed, and it was found that the matter threading the wormhole satisfies the null energy condition, due to the effective stress-energy tensor containing higher order curvature derivatives. Therefore, for case of taking the traceless stress-energy tensor in $f(R,T)$ theory, we do not discuss it in this paper. The other equation of state that we use in this article is $p_t=\beta\rho$ and $p_r=\gamma\rho$. For simplicity and considering the anisotropic property of pressure in WH, we furthermore take model parameter $\gamma=0$ with leaving $\beta$ as the model parameter. Then combining equations (\ref{A6})–(\ref{A8}), we gain the following differential equation
\begin{eqnarray}
4f''_Rr^2(r^2-r^2_0)(11+8\beta)+2r^2_0(2f'_Rr(11+8\beta)-f_R(25+22\beta))=0
\label{A9}
\end{eqnarray}
where $f'_R=\frac{df_R}{dr}$ and $f''_R=\frac{d^2f_R}{dr^2}$. With taking a specific form for function: $b(r)=\frac{r^2_0}{r}$, we gain a solution for the differential equation (\ref{A9}) as,
\begin{eqnarray}
f_R(r)=\frac{W r^\frac{3}{2}\sqrt{r+r_0-H}\sqrt{-r+r_0+H}c_1}{2H^{\frac{1}{2}}\sqrt{r(r-H)}}
\label{A11}
\end{eqnarray}
where
\begin{eqnarray}
W=e^{\frac{\sqrt{6}\sqrt{r^2_0(1+2\beta)}arctan(\frac{-r+H}{r_0})}{r_0\sqrt{11+8\beta}}}
\label{A-1}
\end{eqnarray}
\begin{eqnarray}
H=\sqrt{r-r_0}\sqrt{r+r_0}
\label{A-2}
\end{eqnarray}
where $c_1$ is an integration constant. Thus, the energy density and the pressure of matter can be written as, respectively
\begin{eqnarray}
\rho(r)=-\frac{21 W r^2_0(r^2-r^2_0)^{\frac{3}{4}}(64r^8-144r^6r^2_0+104r^4r^4_0-25r^2r^6_0+r^8_0-64r^7 H +112r^5r^2_0 H -56r^3r^4_0 H +7r r^6_0 H )c_1}{r^{\frac{5}{2}} H \sqrt{r(r-H)}(r+r_0-H)^{\frac{3}{2}}(-r+H)^5(-r+r_0+H)^{\frac{3}{2}}(11+8\beta)(8\pi+\lambda)}
\label{A12}
\end{eqnarray}
\begin{eqnarray}
p_r(r)=0
\label{A13}
\end{eqnarray}
\begin{eqnarray}
p_t(r)=\beta\rho.
\label{A14}
\end{eqnarray}
The classical reconstruction technique is usually utilized to derive $f(R)$ function in the framework of $f(R)$ theory \cite{refr1,refr2,refr3,refr4,refr5}. Here we try to construct an analytical function $f(R)$ for $f(R,T)=f(R)+\lambda T$ theory. Rewriting the trace equation in $f(R,T)$ theory, we gain
\begin{eqnarray}
f(R)=-\frac{1}{2}[(8\pi+\lambda)T+4\lambda P]-\lambda T+\frac{1}{2}Rf_R+\frac{3}{2}\square f_R.
\label{A15}
\end{eqnarray}
According to the relation: $R=\frac{-2b'}{r^2}$ and the traversable condition: $b(r_0)=r_0$, we have $R=\frac{2r^2_0}{r^4}$ and $r=(\frac{2r^2_0}{R})^{\frac{1}{4}}$. Thus utilizing equation (\ref{A15}), we obtain an exact solution,
\begin{eqnarray}
f(R)=\frac{e^{-\frac{\sqrt{6}\sqrt{R_0}\sqrt{\frac{1+2\beta}{R_0}}arctan(\frac{R^{\frac{1}{4}}_0}{R^{\frac{1}{4}}}-K\sqrt{R_0})}{\sqrt{11+8\beta}}}\sqrt{R}(\sqrt{R}-\sqrt{R_0})\sqrt{\frac{2-2R^{\frac{1}{4}}K R^{\frac{1}{4}}_0}{\sqrt{R}\sqrt{R_0}}}\sqrt{\frac{1}{R^{\frac{1}{4}}R^{\frac{1}{4}}_0}}S[4\pi(5+24\beta)+(-8+5\beta)\lambda]c_1}{Y[R^{\frac{1}{4}}(-1+K\sqrt{R_0})-R^{\frac{1}{4}}_0][R^{\frac{1}{4}}(1+K\sqrt{R_0})-R^{\frac{1}{4}}_0][-1+R^{\frac{1}{4}}K R^{\frac{1}{4}}_0]^6(\frac{-1+\frac{\sqrt{R_0}}{\sqrt{R}}}{R_0})^{\frac{1}{4}}R^{\frac{5}{2}}_0(11+8\beta)(8\pi+\lambda)}
\label{A16}
\end{eqnarray}
where
\begin{eqnarray}
S=R^2-25R^{\frac{3}{2}}\sqrt{R_0}+7R^{\frac{7}{4}}K R^{\frac{3}{4}}_0+104RR_0-56R^{\frac{5}{4}}K R^{\frac{5}{4}}_0-144\sqrt{R}R^{\frac{3}{2}}_0+122R^{\frac{3}{4}}K R^{\frac{7}{4}}_0+64R^2_0-64R^{\frac{1}{4}}K R^{\frac{9}{4}}_0
\label{A16-2}
\end{eqnarray}
\begin{eqnarray}
Y=K\sqrt{K+\frac{1}{\sqrt{R_0}}-\frac{1}{R^{\frac{1}{4}}R^{\frac{1}{4}}_0}}\sqrt{-K+\frac{1}{\sqrt{R_0}}+\frac{1}{R^{\frac{1}{4}}R^{\frac{1}{4}}_0}}
\label{A16down-1}
\end{eqnarray}
\begin{eqnarray}
K=\sqrt{\frac{1}{\sqrt{R_0}}+\frac{1}{R^{\frac{1}{4}}R^{\frac{1}{4}}_0}}\sqrt{\frac{-1+\frac{R^{\frac{1}{4}}_0}{R^{\frac{1}{4}}}}{\sqrt{R_0}}}
\label{A-3}
\end{eqnarray}
which is a reconstructed function of $f(R)$ in $f(R,T)$ theory. We plot the picture of $f(R)$ in Fig.\ref{add-ff1}, by taking parameters values: $\beta=0.1$, $c_1=1$ and $\lambda=-30$ (or $\lambda=30$). From Fig.\ref{add-ff1}, we have $f(R)>0$, and it is shown that the effect on the shape of function $f(R)$ is significant for considered variation of values of parameter $\lambda$.

\begin{figure}[ht]
\includegraphics[width=7.5cm, height=5cm]{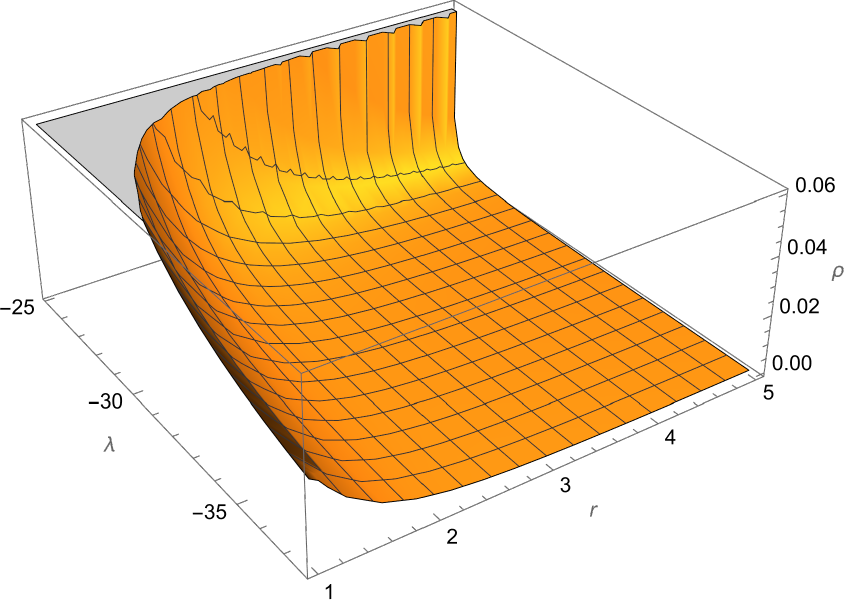}
\includegraphics[width=7.5cm, height=5cm]{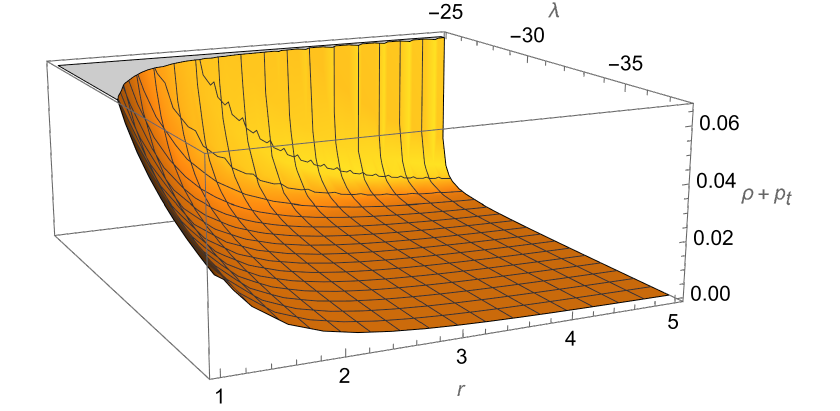}\\
\includegraphics[width=7.5cm, height=5cm]{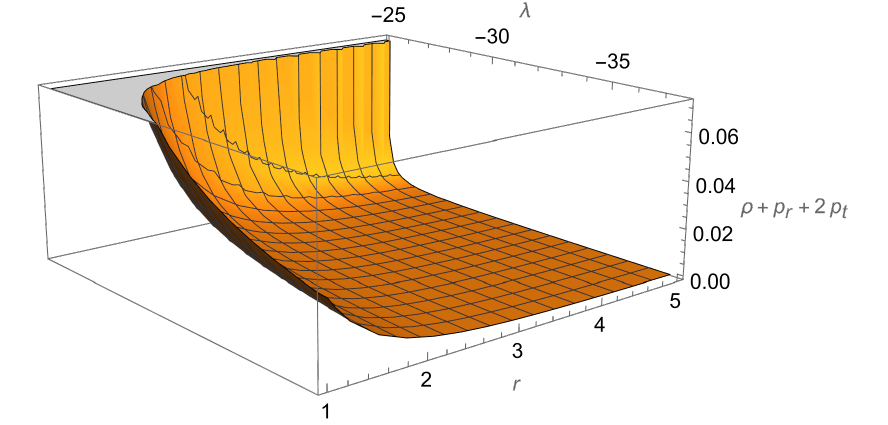}
\includegraphics[width=7.5cm, height=5cm]{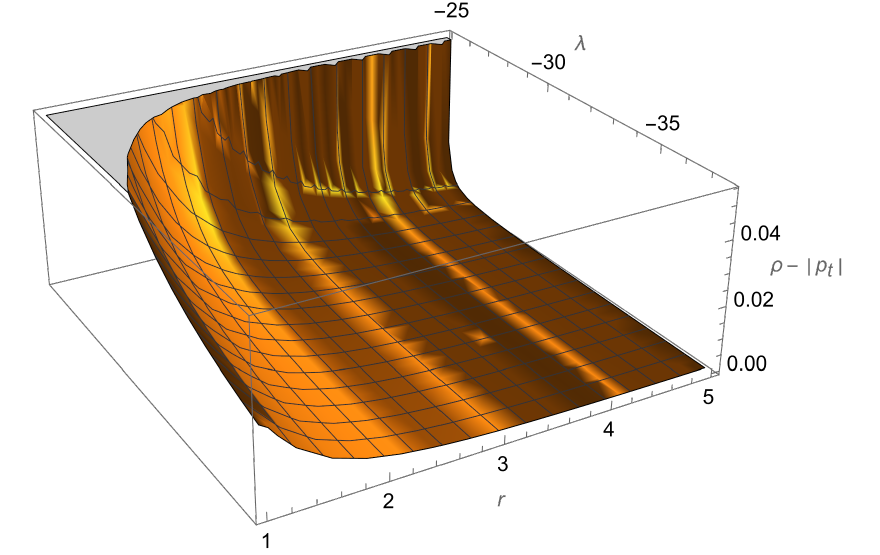}\\
\caption{Pictures of the relevant expressions of energy conditions with respect to parameters $r$ and $\lambda$ in the $f(R,T)$ wormhole, where the model parameter value $\beta=0.1$ is taken.}\label{add-ff2}
\end{figure}

\begin{figure}[ht]
\includegraphics[width=7.5cm, height=5cm]{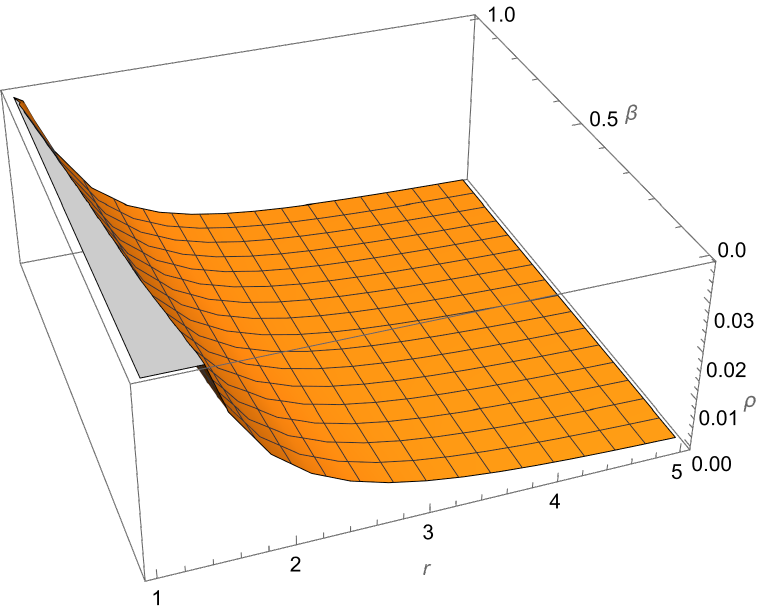}
\includegraphics[width=7.5cm, height=5cm]{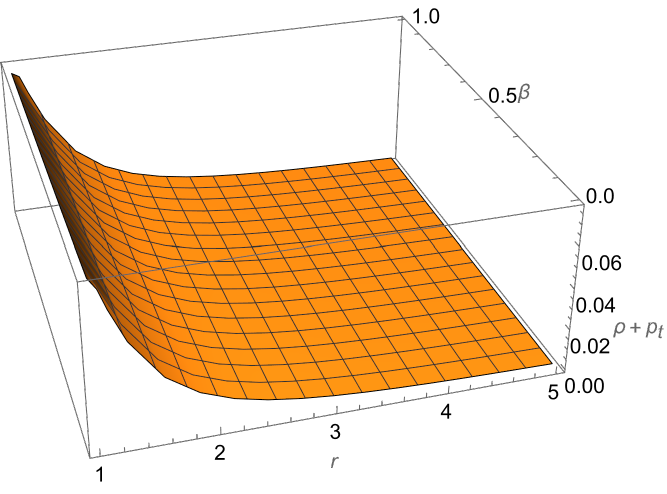}\\
\includegraphics[width=7.5cm, height=5cm]{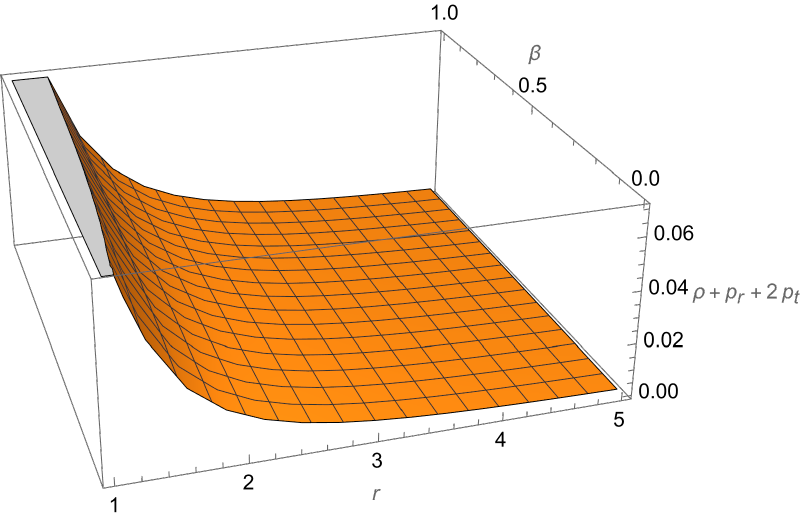}
\includegraphics[width=7.5cm, height=5cm]{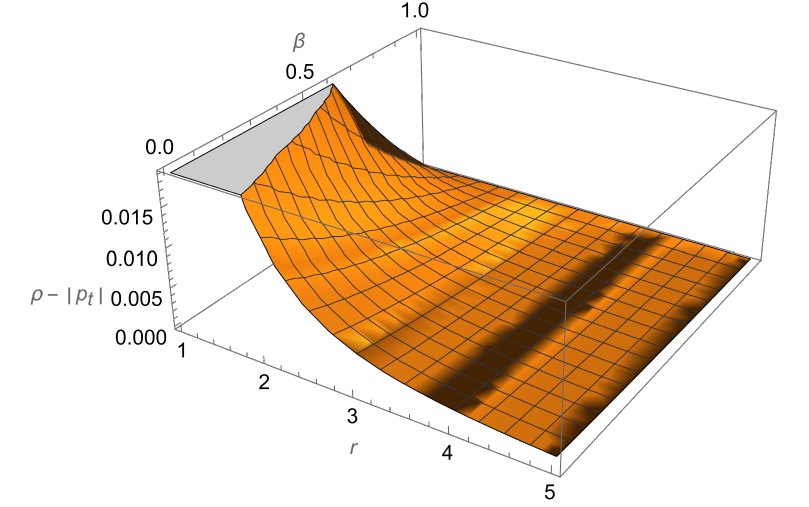}\\
\caption{Pictures of the relevant expressions of energy conditions with respect to parameters $r$ and $\beta$ in the $f(R,T)$ wormhole, where the model parameter value $\lambda=-30$ is taken.}\label{add-ff3}
\end{figure}

In the following analysis, we will discuss the four ECs of matter in this reconstructed $f(R,T)$ wormhole. Given that the radial pressure $p_r=0$ in this work,  we only need to draw four images, i.e. the three-dimensional pictures: $\rho$, $\rho+p_t$, $\rho+p_r+2p_t$, and $\rho-\left| p_t \right |$ with respect to $r$ (please see Fig.\ref{add-ff2} and Fig.\ref{add-ff3}), for inspecting the ECs of matter in this $f(R,T)$ traversable wormhole.  We also investigated the impact of different model parameter values on material energy conditions. Concretely, in Fig.\ref{add-ff2} we limited $-12\pi<\lambda<-8\pi$ and $\beta=0.1$;  in Fig.\ref{add-ff3} we selected model-parameter values with $\lambda=-30$ and $0<\beta<1$. The radius of wormhole throat $r_0=1$ and the integration constant $c_1=1$ have been fixed at prior.

From Fig.\ref{add-ff2} and Fig.\ref{add-ff3}, we can see that all four ECs (NEC, WEC, SEC and DEC) of matter in the traversable wormholes are valid in this reconstructed $f(R,T)$ model, i.e we provide a wormhole solution without introducing the exotic matter and special matter in the modified $f(R,T)$ gravitational theory.

\section{$\text{Conclusion}$}

In this work, we applied the theoretical model $f(R,T)=R+\alpha R^2+\lambda T$ to study the static traversable WH. In this model, the model parameter $\alpha>0$ implies cosmic inflation, and another parameter must satisfy $\lambda<-8\pi$ to ensure positive energy density. Further research revealed that we can obtain traversable WH geometry without introducing exotic matter violating the NEC in the $f(R,T)$ theory. The violation of the DEC could be thought to be related to quantum fluctuations or imply the presence of special matter violating this EC in the wormhole. Our study also demonstates that, compared to GR, the geometric term $\alpha R^2$ has little effect on the wormhole matter compositions and properties in the $f(R,T)=R+\alpha R^2+\lambda T$ model. While the injection of the matter-geometry coupling term $\lambda T$ can resolve the issues of wormhole matter violating the NEC, WEC and SEC in GR. In addition, we investigated the dependence of the validity of the NEC on model parameters and quantified the WH matter components by using the ``volume integral quantifier". Lastly, based on the modified Tolman-Openheimer-Volkov equation, we analyzed the stability of traversable wormholes in this theory.

It is very interesting if one can obtain a WH solution without introducing the exotic matter and special matter in modified gravitational theory. We utilize the classical reconstruction technique to derive wormhole solution in $f(R,T)$ theory and discuss the energy conditions of matter in this solution. It is found that all four ECs (NEC, WEC, SEC and DEC) of matter in the traversable wormholes are valid in this reconstructed $f(R,T)$ model, i.e we provide a wormhole solution without introducing the exotic matter and special matter in the modified $f(R,T)$ gravitational theory. 

\textbf{\ Acknowledgments }
 The research work is supported by   the National Natural Science Foundation of China (12175095,12075109 and 11865012), and supported by  LiaoNing Revitalization Talents Program (XLYC2007047).
\textbf{\ 'Data availability' statement }
 My manuscript has no associated data. Data sharing is not applicable to this article as no new data were created or analyzed in this study.

\end{document}